\documentclass[copyright,creativecommons]{eptcs}

\providecommand{\event}{ACL2-2023}

\usepackage{alltt}
\usepackage{amssymb}
\usepackage{caption}
\usepackage{enumitem}
\usepackage{eucal}
\usepackage[T1]{fontenc}
\usepackage{inconsolata}
\usepackage{mathtools}
\usepackage[inference]{semantic}
\usepackage{subcaption}
\usepackage{tikz}
\usepackage{underscore}
\usepackage{url}
\usepackage{wrapfig}
\usepackage{xcolor}

\newcommand{\secref}[1]{Section \ref{#1}}
\newcommand{\secrefII}[2]{Sections \ref{#1} and \ref{#2}}
\newcommand{\figref}[1]{Figure \ref{#1}}

\newcommand{\citecode}[2]
  {\cite[\href{https://github.com/acl2/acl2/tree/master/#1}{\texttt{#2}}]
        {acl2-code}}
\newcommand{\citeman}[2]
  {\cite[\href{http://acl2.org/manual?topic=#1}{\texttt{#2}}]
        {acl2-manual}}

\newcommand{\code}[1]{\texttt{#1}} 
\definecolor{commentgray}{gray}{0.4}
\newenvironment{bcode} 
 {\begin{quote}\small\begin{alltt}}
 {\end{alltt}\end{quote}}

 \newcommand{\logAnd}[2]{#1\wedge#2}
 \newcommand{\logImp}[2]{#1\Longrightarrow#2}

\newcommand{\logAll}[3][]{\forall#2.\:#1#3}
\newcommand{\logEx}[3][]{\exists#2.\:#1#3}

\newcommand{\set}[1]{\{#1\}}
\newcommand{\setI}[1]{\set{#1}}

\newcommand{\setFT}[2]{\set{#1,\ldots,#2}}
\newcommand{\setST}[2]{\set{#1\mid#2}}

\newcommand{\setIn}[3][]{#2#1\in#1#3}

\newcommand{\setSub}[2]{#1\subseteq#2}

\newcommand{\setSup}[2]{#1\supseteq#2}

\newcommand{\setUni}[2]{#1\cup#2}

\newcommand{\isfun}[3]{#1:#2\rightarrow#3}
\newcommand{\isrel}[2]{#1\subseteq#2}

\newcommand{\tuple}[1]{\langle#1\rangle}
\newcommand{\tupleO}{\tuple{\,}}

\newcommand{\tupleIII}[3]{\tuple{#1,#2,#3}}
\newcommand{\tupleIV}[4]{\tuple{#1,#2,#3,#4}}
\newcommand{\tupleFT}[2]{\tuple{#1,\ldots,#2}}

\newcommand{\maplet}[2]{#1\mapsto#2}

\newcommand{\zksecret}{x}
\newcommand{\zkpred}{P}
\newcommand{\zkpredlo}{\zkpred_\mathrm{L}}
\newcommand{\zkpredhi}{\zkpred_\mathrm{H}}

\newcommand{\modulo}[2]{#1\ \mathsf{mod}\ #2}
\newcommand{\modulop}[1]{\modulo{#1}{p}}

\newcommand{\Field}{\mathbb{F}}
\newcommand{\Fieldof}[1]{\Field_{#1}}
\newcommand{\Fieldp}{\Fieldof{p}}

\newcommand{\faddSYM}[1]{\oplus_{#1}}
\newcommand{\fsubSYM}[1]{\ominus_{#1}}
\newcommand{\fmulSYM}[1]{\otimes_{#1}}
\newcommand{\fdivSYM}[1]{\oslash_{#1}}

\newcommand{\fadd}[3]{#1 \faddSYM{#3} #2}
\newcommand{\fsub}[3]{#1 \fsubSYM{#3} #2}
\newcommand{\fmul}[3]{#1 \fmulSYM{#3} #2}
\newcommand{\fdiv}[3]{#1 \fdivSYM{#3} #2}

\newcommand{\faddp}[2]{\fadd{#1}{#2}{p}}
\newcommand{\fsubp}[2]{\fsub{#1}{#2}{p}}
\newcommand{\fmulp}[2]{\fmul{#1}{#2}{p}}
\newcommand{\fdivp}[2]{\fdiv{#1}{#2}{p}}

\newcommand{\rrel}{R}
\newcommand{\rrelx}{\widetilde{\rrel}}

\newcommand{\extvar}{\phi}
\newcommand{\intvar}{\phi'}

\newcommand{\spec}{S}

\newcommand{\compfun}{f}
\newcommand{\compfunf}{\widehat{\compfun}}

\newcommand{\comperr}{\mathcal{E}}

\newcommand{\Indata}{I}
\newcommand{\Outdata}{O}

\newcommand{\encode}{e}
\newcommand{\encodein}{\encode^\mathrm{I}}
\newcommand{\encodeout}{\encode^\mathrm{O}}

\newenvironment{r1cs}
 {\begin{array}{rlcl}}
 {\end{array}}
\newcommand{\rics}[3]{(#1) \!\! & \!\! (#2) & \!\!\!\! = \!\!\!\! & (#3)}
\newcommand{\ricsdots}{&&\vdots&}

\newcommand{\pfcsNonterm}[1]{\mathsf{#1}} 
\newcommand{\pfcsInt}{\pfcsNonterm{I}} 
\newcommand{\pfcsNam}{\pfcsNonterm{N}} 
\newcommand{\pfcsExp}{\pfcsNonterm{E}} 
\newcommand{\pfcsCon}{\pfcsNonterm{C}} 
\newcommand{\pfcsRel}{\pfcsNonterm{R}} 
\newcommand{\pfcsTerm}[1]{\mbox{\texttt{#1}}} 
\newcommand{\pfcsEq}{\!\!\!\!::=\!\!\!\!}
\newcommand{\pfcsAlt}{\ \mid\ } 

\newcommand{\pfcsexp}{e} 
\newcommand{\pfcscon}{c} 
\newcommand{\pfcsrel}{r} 
\newcommand{\pfcsvar}{v} 

\newcommand{\pfcsSat}[4][]{#3#1\vdash_{#2}#1#4}
\newcommand{\pfcsAsg}{\alpha} 
\newcommand{\pfcsAsgx}{\pfcsAsg'} 
\newcommand{\pfcsRels}{\rho} 
\newcommand{\pfcsVal}{\phi} 

\begin{document}


\title{Formal Verification of Zero-Knowledge Circuits}

\author{Alessandro Coglio
        \quad\quad
        Eric McCarthy
        \quad\quad
        Eric W. Smith
        \institute{Kestrel Institute \quad \url{https://kestrel.edu} \\
                   Aleo Systems Inc. \quad \url{https://aleo.org}}}

\def\titlerunning{Formal Verification of Zero-Knowledge Circuits}
\def\authorrunning{A. Coglio, E. McCarthy, E. Smith}

\maketitle

\begin{abstract}
Zero-knowledge circuits are sets of equality constraints
over arithmetic expressions interpreted in a prime field;
they are used to encode computations in cryptographic zero-knowledge proofs.
We make the following contributions to
the problem of ensuring that a circuit correctly encodes a computation:
a formal framework for circuit correctness;
an ACL2 library for prime fields;
an ACL2 model of the existing R1CS (Rank-1 Constraint Systems) formalism
to represent circuits,
along with ACL2 and Axe tools to verify circuits of this form;
a novel PFCS (Prime Field Constraint Systems) formalism
to represent hierarchically structured circuits,
along with an ACL2 model of it
and ACL2 tools to verify circuits of this form
in a compositional and scalable way;
verification of circuits, ranging from simple to complex;
and discovery of bugs and optimizations in existing zero-knowledge systems.
\end{abstract}


\section{Introduction}
\label{intro}

In cryptography, a \emph{zero-knowledge proof} is a method by which
a \emph{prover} can convince a \emph{verifier}
that they know a secret $\zksecret$
that satisfies a computable predicate $\zkpred$,
without revealing $\zksecret$ and without involving third parties
\cite{zkproofs-conference,zkproofs-noninteractive}.
Spurred by recent advances that have greatly improved their efficiency
\cite{snarks,groth16,bulletproofs,zkstarks,halo},
zero-knowledge proofs are finding increasingly wide application,
particularly in the blockchain world
\cite{eth-zk-rollups,zerocoin,zerocash,zcash,zexe,aleo},
holding promise to rebalance privacy on the Internet
\cite{aleo-rebalance,web3-wired}.

While most of the technical details of zero-knowledge proofs
are irrelevant to this paper,
the one crucial fact is that
the predicate $\zkpred$ must be expressed as a \emph{zero-knowledge circuit},
which can be defined%
\footnote{There seems to be no universal definition of
zero-knowledge circuits and of some related notions in the literature.}
as a set of equality constraints over integer variables
where the only operations are
addition and multiplication modulo a large prime number.
This is a low-level representation $\zkpredlo$ of $\zkpred$,
at odds with the need for $\zkpred$
to be clearly understood by both prover and verifier,
who we presume would understand
a higher-level representation $\zkpredhi$ of $\zkpred$,
e.g.\ expressed in a conventional programming language.
Unless $\zkpredlo$ and $\zkpredhi$ denote the same $\zkpred$,
the zero-knowledge proof may not quite prove what is expected.

This leads to the mathematically well-defined problem
of formally proving that a zero-knowledge circuit
correctly represents a higher-level description.
Note the difference between formal proofs,
which provide logic-based unconditional evidence of mathematical assertions,
and zero-knowledge proofs,
which provide cryptography-based statistically overwhelming evidence
of computational assertions.
Besides formal proofs about zero-knowledge proofs,
which is the topic of this paper,
one could imagine doing zero-knowledge proofs of formal proofs
(i.e.\ prove a theorem without revealing the proof,
which may have interesting applications),
but we have not explored that yet.
Given the above characterization of the problem
in terms of zero-knowledge circuits,
the zero-knowledge proof aspect is largely irrelevant here;
the unqualified `proof' and similar words in the rest of this paper
have the familiar meaning.

This paper describes our endeavors, in the course of various projects,
to tackle the zero-knowledge circuit verification problem,
using ACL2 \cite{acl2-www} and tools built on it.
Our contributions are:
\begin{enumerate}[nosep,label=(\alph*)]
\item\label{contrib-frame}
A general formal framework for zero-knowledge circuit correctness,
i.e.\ that $\zkpredlo \Longleftrightarrow \zkpredhi$.
\item\label{contrib-pfield}
A library of rules to reason about \emph{prime fields}---%
the arithmetic basis for zero-knowledge circuits.
\item\label{contrib-r1cs-model}
A formal model of \emph{Rank-1 Constraint Systems} (\emph{R1CS}),
an existing formalism commonly used to represent zero-knowledge circuits.
\item\label{contrib-r1cs-tools}
Rules and tools to verify R1CS circuits,
including a new specialized version of Axe \citeman{ACL2____AXE}{axe}.
\item\label{contrib-pfcs-model}
A formal model of \emph{Prime Field Constraint Systems} (\emph{PFCS}),
a novel formalism developed by us that generalizes R1CS
with richer forms of constraints and with hierarchical structure.
\item\label{contrib-pfcs-tools}
Rules and tools to verify PFCS circuits,
in a compositional and scalable way.
\item\label{contrib-proof-examples}
Verification of zero-knowledge circuits, ranging from simple to complex,
in R1CS and PFCS form.
\item\label{contrib-bugs-and-optimizations}
Discovery of two bugs and several optimizations
in a zero-knowledge circuit construction library.
\end{enumerate}

\secref{back} provides the necessary background on zero-knowledge circuits.
\secref{frame} describes contribution \ref{contrib-frame}.
\secref{pfield} describes contribution \ref{contrib-pfield}.
\secref{r1cs} describes contributions
\ref{contrib-r1cs-model},
\ref{contrib-r1cs-tools},
\ref{contrib-proof-examples}, and
\ref{contrib-bugs-and-optimizations}.
\secref{pfcs} describes contributions
\ref{contrib-pfcs-model},
\ref{contrib-pfcs-tools}, and
\ref{contrib-proof-examples}.
Related work is discussed in \secref{related}.
Future work is outlined in \secref{future}.
Some conclusions are drawn in \secref{concl}.

\section{Background}
\label{back}

\noindent
A \emph{prime field} is a set $\Fieldp = \setFT{0}{p-1}$,
consisting of the natural numbers below $p$,
where $p$ is a prime number.
The \emph{arithmetic operations} on $\Fieldp$ are:
\\
\begin{tabular}{ll}
\emph{addition}:       & $\faddp{x}{y} = \modulop{(x + y)}$ \\
\emph{subtraction}:    & $\fsubp{x}{y} = \modulop{(x - y)}$ \\
\emph{multiplication}: & $\fmulp{x}{y} = \modulop{(x \times y)}$ \\
\emph{division}:       & $\fdivp{x}{y} = z$,
                         where $x = \fmulp{y}{z}$,
                         if $y \neq 0$
\end{tabular}
\\
That is, all the operations are modular versions of the ones on the integers,
except that the division of $x$ by $y$
yields the unique $z$ (which always exists)
that yields $x$ when multiplied by $y$,
provided that $y \neq 0$.
We may denote the prime field arithmetic operations
with the same symbols as the integer arithmetic operations,
i.e.\ $+$, $-$, $\times$, $/$.
We may also omit the multiplication symbol altogether,
e.g.\ $(x + 1) (y - 1)$ may stand for $(x + 1) \times (y - 1)$,
which in turn may stand for $\fmulp{(\faddp{x}{1})}{(\fsubp{y}{1})}$.
We may also just write $\Field$, leaving $p$ implicit.
Context should always disambiguate these commonly used abbreviations.

A \emph{zero-knowledge circuit} is
a set of \emph{constraints}
that are equalities between \emph{expressions}
built out of
\emph{variables},
\emph{constants},
\emph{additions}, and
\emph{multiplications},
all interpreted in $\Field$.
By designating certain variables as \emph{inputs} and \emph{outputs},
the constraints can represent a computation of outputs from inputs.
Zero-knowledge circuits generalize \emph{arithmetic circuits},
which are like boolean circuits,
except that wires carry integers instead of booleans,
and gates perform arithmetic operations instead of boolean ones.

For reasons that depend on the details of zero-knowledge proofs,
such constraints must be written in specific forms \cite{buterin-qap}.
A popular formalism is
\emph{Rank-1 Constraint Systems} (\emph{R1CS}),
whose constraints have the form
$
(a_0 + a_1 x_1 + \cdots + a_n x_n)
\
(b_0 + b_1 y_1 + \cdots + b_m y_m)
=
(c_0 + c_1 z_1 + \cdots + c_l z_l)
$,
where $n,m,l \geq 0$,
each $a_i, b_j, c_k$ is a \emph{coefficient} in $\Field$,
and each $x_i, y_j, z_k$ is a \emph{variable} ranging over $\Field$.
That is, an R1CS constraint is an equality between
the product of two polynomials and a polynomial,
each polynomial having zero or more variables with exponent 1,
i.e.\ a \emph{linear combination}.
An R1CS circuit is a set of these constraints.
Literature definitions of R1CS are usually in terms of vectors and matrices;
the definition just given here is more like an abstract syntax of R1CS.

\begin{wrapfigure}[6]{r}{3.2in}
\centering
\vspace*{-0.3in}
\[
\begin{r1cs}
\rics{w}{x-y}{z-y}
\end{r1cs}
\quad\quad
z :=
\begin{cases}
x & \mbox{if } w = 1 \\
y & \mbox{if } w = 0
\end{cases}
\]
\vspace*{-0.1in}
\caption{An `if-then-else' conditional.}
\label{fig:r1cs-cond}
\end{wrapfigure}

For example,
if $w$ is \emph{boolean}, i.e.\ either 1 or 0,
the circuit in the left part of \figref{fig:r1cs-cond} represents
the computation in the right part,
which sets $z$ to $x$ or $y$ based on whether $w$ is 1 or 0.
If $w = 0$, the left side of the constraint is 0 and thus $z = y$;
if $w = 1$, the $-y$ cancels and thus $z = x$.

\begin{wrapfigure}[6]{r}{3.5in}
\centering
\vspace*{-0.2in}
\[
\begin{r1cs}
\rics{u-v}{s}{1-w} \\
\rics{u-v}{w}{0}
\end{r1cs}
\quad\quad
w :=
\begin{cases}
1 & \mbox{if } u = v \\
0 & \mbox{if } u \neq v
\end{cases}
\]
\vspace*{-0.1in}
\caption{An equality test.}
\label{fig:r1cs-eq}
\end{wrapfigure}

As another example,
the circuit in the left part of \figref{fig:r1cs-eq} represents
the computation in the right part,
which sets $w$ to 1 or 0 based on whether $u = v$ or not.
If $u = v$, the first constraint makes $w = 1$,
and the second constraint is satisfied because $0 \times 1 = 0$.
If $u \neq v$, the second constraint makes $w = 0$,
and the first constraint is satisfied by $s = 1 / (u - v)$.

These and other examples can be found in the literature,
e.g.\ \cite{zk0x04-r1cs-notes} and \cite[Appendix A]{zcash-spec}.
Circuits vary in size and complexity.
Even the ones in \figref{fig:r1cs-cond} and \figref{fig:r1cs-eq}
require a little thought to understand.

Larger circuits are built from smaller ones
by joining their constraints and sharing some variables.
For instance,
combining \figref{fig:r1cs-cond} and \figref{fig:r1cs-eq}
yields a circuit that represents
the computation that sets $z$ to $x$ or $y$ based on whether $u = v$ or not;
the variable $w$ is shared,
with \figref{fig:r1cs-eq} guaranteeing that it is boolean
as assumed in \figref{fig:r1cs-cond}.
In this kind of hierarchical construction,
a \emph{gadget} is a circuit with a well-defined purpose,
usable as a component of larger gadgets,
and possibly made of smaller gadgets.
The zero-knowledge circuit $\zkpredlo$ in \secref{intro}
is a top-level gadget;
it represents the computation,
described in some high-level way $\zkpredhi$,
of the predicate $\zkpred$ on the secret input $\zksecret$.

While all the variables in the gadget in \figref{fig:r1cs-cond}
are involved in the represented computation,
the variable $s$ in the gadget in \figref{fig:r1cs-eq} is not.
It is an \emph{internal} variable,
while the other ones are \emph{external} variables;
the latter are divided into \emph{input} and \emph{output} variables
according to the represented computation.
When the two gadgets are combined as just described,
the shared external variable $w$
becomes internal to the combined gadget.
The distinction between external and internal variables,
and between input and output variables,
is not captured in the R1CS formalism,
but it is arguably implicit in the notion of gadget.
In general, internal variables cannot be avoided in gadgets;
attempts to eliminate them often result in subtly non-equivalent constraints
that fail to adequately represent the intended computation.

Although direct support is limited to
$\Field$ as a data type
and (field) addition and multiplication as operations,
R1CS circuits are at least as expressive as boolean circuits:
if $x$ and $y$ are boolean variables (like $w$ earlier),
the constraint $(x) (y) = (1 - z)$
represents a `nand' gate with output $z$
(and similarly simple constraints represent other logical gates);
and higher-level data types can be always encoded as bits.
But more efficient representations (fewer variable and constraints)
are often possible.

\begin{wrapfigure}[8]{r}{3.2in}
\centering
\vspace*{-0.2in}
\[
\begin{r1cs}
\rics
 {z_0}
 {1 - z_0}
 {0} \\
\ricsdots \\
\rics
 {z_n}
 {1 - z_n}
 {0} \\
\rics
 {\sum_{i=0}^{n} 2^i z_i}
 {1}
 {\sum_{i=0}^{n-1} 2^i x_i + \sum_{i=0}^{n-1} 2^i y_i}
\end{r1cs}
\]
\vspace*{-0.1in}
\caption{An unsigned $n$-bit integer addition.}
\label{fig:r1cs-add}
\end{wrapfigure}

For example,
two unsigned $n$-bit integers,
encoded as the bits $x_0,\ldots,x_{n-1}$ and $y_0,\ldots,y_{n-1}$
in little endian order,
can be added via the gadget in \figref{fig:r1cs-add}.
The first $n+1$ constraints force $z_0,\ldots,z_n$ to be boolean.
The last constraint forces them to be the bits of the sum,
in little endian order, where $z_n$ is the carry.
This assumes that the prime $p$ has at least $n+2$ bits,
so that the field operations do not wrap around $p$;
a typical $p$ has about 250 bits,
sufficient for fairly large integers.

R1CS circuits are normally constructed programmatically
using libraries \cite{snarkvm,ark-r1cs-std,bellman,librustzcash,libsnark}
that provide facilities to build gadgets hierarchically.
These libraries are invoked directly,
by programs written to build specific circuits,
or indirectly,
by compilers of higher-level languages to R1CS
\cite{leo,noir,circom,lurk-www,zksecrec,coda}.
As these libraries are invoked,
they generate growing sequences of the R1CS constraints that form the gadgets.%
\footnote{The final sequence
consists of the constraints for the predicate $\zkpred$ in \secref{intro},
and is part of the zero-knowledge proof.}
Separate instances of the same gadget have different variables,
which are typically generated via monotonically increasing indices.
The gadgets' hierarchical structure,
reflected in both the static organization and the dynamic execution
of the libraries,
is lost in the generated flat sequence of constraints;
this is not an issue for zero-knowledge proofs,
but it can be for formal proofs, as elaborated later.

\section{Formal Framework}
\label{frame}

An R1CS circuit, along with an ordering of the $r$ variables that occur in it,
determines a relation $\isrel{\rrel}{\Field^r}$
consisting of the $r$-tuples that satisfy all the constraints,
when assigned element-wise to the variables.
If additionally the variables are
partitioned into $q$ external ones and $r-q$ internal ones
(in the sense of \secref{back}),
and ordered so that the former precede the latter,
a relation $\isrel{\rrelx}{\Field^q}$ is also determined,
defined as
$\rrelx =
 \setST
  {\tupleFT{\extvar_1}{\extvar_q}}
  {\logEx
    {\tupleFT{\intvar_1}{\intvar_{r-q}}}
    {\rrel(\extvar_1,\ldots,\extvar_q,\intvar_1,\ldots,\intvar_{r-q})}}$,
i.e.\ consisting of the $q$-tuples that satisfy all the constraints,
when assigned element-wise to the external variables,
for some $(r-q)$-tuples assigned element-wise to the internal variables.
The tuples are \emph{solutions} of the constraints.
The informal notion of \emph{gadget} described in \secref{back}
can be more precisely defined as an R1CS circuit accompanied by
an ordering and designation of its variables as just described;
$\rrelx$ is the semantics of the gadget.
For example, for the gadget in \figref{fig:r1cs-eq},
$\rrel =
 \setST
  {\tupleIV{u}{v}{w}{s}}
  {\logAnd
    {(u-v) \: s = 1-w\:}
    {\:(u-v) \: w = 0}}$
and
$\rrelx =
 \setST
  {\tupleIII{u}{v}{w}}
  {\logEx{s}{R(u,v,w,s)}}$.

Given this semantic characterization,
it is natural
to use a relation $\isrel{\spec}{\Field^q}$
as \emph{specification} of the gadget,
and to express \emph{correctness} of the gadget as $\rrelx = \spec$.
The specification $\spec$ may be defined
in whichever high-level way that is convenient (more on this later),
but in any case it denotes the set of $q$-tuples
that must be the solutions of the gadget.
Correctness consists of
\emph{soundness} $\setSub{\rrelx}{\spec}$
(i.e.\ every solution of the gadget satisfies the specification)
and
\emph{completeness} $\setSub{\spec}{\rrelx}$
(i.e.\ everything satisfying the specification is a solution of the gadget).
To prove soundness and completeness,
the definition of $\rrelx$ must be expanded,
to expose the constraints that define $\rrel$.
To prove soundness,
the existential quantification over the antecedent
can be turned into a universal quantification over the implication,
leading to the quantifier-free formula
$\logImp
  {\rrel(\extvar_1,\ldots,\extvar_q,\intvar_1,\ldots,\intvar_{r-q})}
  {\spec(\extvar_1,\ldots,\extvar_q)}$.
To prove completeness, no such move is possible:
the formula
$\logImp
  {\spec(\extvar_1,\ldots,\extvar_q)}
  {\logEx
    {\tupleFT{\intvar_1}{\intvar_{r-q}}}
    {\rrel(\extvar_1,\ldots,\extvar_q,\intvar_1,\ldots,\intvar_{r-q})}}$
demands dealing with the existential quantification explicitly,
typically by exhibiting witnesses for
the internal variables $\intvar_1,\ldots,\intvar_{r-q}$.

When a gadget represents a computation (as in \secref{back}),
the specification $\spec$ must specify the computation.
For this, a \emph{computation} is modeled as a function
$\isfun
  {\compfun}
  {\Indata_1\times\cdots\times\Indata_n}
  {\setUni
    {(\Outdata_1\times\cdots\times\Outdata_m)}
    {\setI{\comperr}}}$
from $n \geq 0$ inputs to $m \geq 0$ outputs or to a distinct error $\comperr$.
The case $n = 0$ is uninteresting but unproblematic.
The case $m = 0$ models assertion-like computations,
e.g.\ for each gadget $(z_i) (1 - z_i) = (0)$
that checks whether $z_i$ is a bit,
used in \figref{fig:r1cs-add}:
the computation represented by the gadget returns
the empty tuple of outputs $\tupleO$ if $z_i$ is boolean,
or $\comperr$ otherwise.
The function $\compfun$ always models a deterministic computation,
which is appropriate for zero-knowledge applications.%
\footnote{While zero-knowledge proofs themselves
involve non-deterministic computations,
normally they (probabilistically) prove facts about deterministic computations.}
The function $\compfun$ only captures the computation's input/output behavior,
not other aspects of its execution;
this is consistent with the fact that R1CS constraints
only express relations among variables.
For example, for the gadget in \figref{fig:r1cs-eq},
$\Indata_1 = \Indata_2 = \Outdata_1 = \Field$,
$\compfun(u,v) = 1$ if $u = v$,
$\compfun(u,v) = 0$ if $u \neq v$,
and thus $\compfun(u,v) \neq \comperr$ always.

To represent $\compfun$ in R1CS form,
its inputs and outputs must be represented
as field elements,
via injective encoding functions
$\isfun{\encodein_i}{\Indata_i}{\Field^{n_i}}$,
where ${\encodein_i}$ maps each input ${\setIn{x_i}{\Indata_i}}$
to some number ${n_i}$ of field elements,
and
$\isfun{\encodeout_j}{\Outdata_j}{\Field^{m_j}}$,
where ${\encodeout_j}$ maps each output ${\setIn{y_j}{\Outdata_j}}$
to some number ${m_j}$ of field elements.
This leads to a computation on encoded inputs and outputs
$\isfun{\compfunf}{\Field^N}{\setUni{\Field^M}{\setI{\comperr}}}$,
with $N = \sum_i n_i$ and $M = \sum_j m_j$,
defined as follows:
(1) if
$\compfun(x_1,\ldots,x_n) = \tupleFT{y_1}{y_m}$
then
$\compfunf(\encodein_1(x_1),\ldots,\encodein_n(x_n)) =
 \tupleFT{\encodeout_1(y_1)}{\encodeout_m(y_m)}$;
(2) if
$\compfun(x_1,\ldots,x_n) = \comperr$
then
$\compfunf(\encodein_1(x_1),\ldots,\encodein_n(x_n)) = \comperr$;
and (3)
$\compfunf$ returns $\comperr$ outside the range of the input encodings.
For example, for the gadget in \figref{fig:r1cs-eq},
$\encodein_1 = \encodein_2 = \encodeout_1 = \mathit{id}$ (identity)
and $\compfunf = \compfun$.

The computation $\compfunf$ is represented by a gadget
with $q = N + M$ external variables for the inputs and outputs.
The specification of $\compfunf$ is
$\spec(x_1,\ldots,x_N,y_1,\ldots,y_M) =
 [\compfunf(x_1,\ldots,x_N) = \tupleFT{y_1}{y_M}]$.
Thus,
soundness means that
every solution of the gadget corresponds to
a non-erroneous instance of the computation,
and completeness means that
every non-erroneous instance of the computation corresponds to
a solution of the gadget.
Soundness alone is not sufficient for correctly representing a computation:
a gadget without solutions is trivially sound;
completeness ensures that there is a solution
for every input for which the computation is not erroneous.
For example, for the gadget in \figref{fig:r1cs-eq},
$\spec =
 \setST
  {\tupleIII{u}{v}{w}}
  {\compfun(u,v) = w}$.

For a top-level gadget $\zkpredlo$
that represents $\zkpred$ in a zero-knowledge proof
(see \secref{intro} and \secref{back}),
whose formal semantics is a relation $\rrelx_\zkpred$ as above,
the specification $\spec_\zkpred$ is derived
from the high-level description $\zkpredhi$ of $\zkpred$.
If $\zkpredhi$ is a program in a higher-level language
\cite{leo,noir,circom,lurk-www,zksecrec,coda},
a function $\compfun_\zkpred$ that denotes the execution of the program
is formally defined, based on a formalization of the language,
and a specification relation $\spec_\zkpred$ is derived from it as above.
If $\zkpredhi$ is a description of a fixed application-specific computation
(e.g.\ Zcash shielded transactions \cite{zcash}),
$\compfun_\zkpred$ is defined by formalizing that description,
and $\spec_\zkpred$ is derived from it as above.
For sub-gadgets of $\zkpredlo$,
specifications may be written in any formal form that is convenient;
these specifications play a role in the formal verification of $\zkpredlo$
(see \secref{compver} and \secref{future}),
but they are not directly exposed to
the zero-knowledge prover and verifier mentioned in \secref{intro}.
The understandability of $\zkpredhi$ by prover and verifier,
mentioned in \secref{intro},
as with all complex technologies,
boils down to
trusting authoritative high-level informal descriptions
for users from the general public,
analyzing the aforementioned programs
for users who are also software developers,
and examining the formalizations and theorems
for users who are also formal verification specialists.

\section{Prime Fields}
\label{pfield}

Starting with the recognizer of prime numbers \code{primep} from
\citecode{books/projects/numbers}{[books]/projects/numbers},
the \emph{prime fields library}
\citeman{PFIELD____PRIME-FIELDS}{prime-fields}
introduces a recognizer \code{fep} of field elements,
and functions
\code{add}, \code{sub}, \code{mul}, \code{div}, \code{neg}, \code{inv},
\code{pow}, and \code{minus1}
for field operations.
These are all parameterized over a prime \code{p},
e.g.\ \code{(fep x p)} checks if
\code{x} is in $\Fieldof{\code{p}}$,
and \code{(add x y p)}
returns $\fadd{\code{x}}{\code{y}}{\code{p}}$
(see \secref{back}).

The recognizer and operations are executable.
The multiplicative inverse \code{inv} is calculated via \code{pow},
according to the known equation
$\fdiv{1}{x}{p} = \modulo{x^{p-2}}{p}$;
we prove that this definition indeed yields the multiplicative inverse.
The definition of \code{pow} is an \code{mbe}
whose \code{:logic} is recursively repeated multiplication
and whose \code{:exec} is fast modular exponentiation
\code{mod-expt-fast} from \citeman{ACL2____ARITHMETIC-3}{arithmetic-3}.

The library provides basic theorems,
such as all the standard field axioms (e.g.\ commutativity of addition);
these theorems often suffice for relatively simple reasoning.
The library also provides collections of rules
that realize certain normalization strategies,
useful for more elaborate reasoning.

\section{Rank-1 Constraint Systems}
\label{r1cs}

\subsection{Model}
\label{r1cs-model}

Based on the prime fields library described in \secref{pfield},
the \emph{R1CS library} \citeman{R1CS____R1CS}{r1cs}
provides an ACL2 model of R1CS.
It follows the nomenclature of literature definitions of R1CS,
which are in terms of vectors and matrices.
The model consists of
a \emph{dense} formulation,
where linear combinations have monomials for all the involved variables
(many with zero coefficients),
and a \emph{sparse} formulation,
where linear combinations may omit monomials with zero coefficients.
The dense formulation is of intellectual interest
but impractical for verification;
the rest of this paper focuses on the sparse formulation.

The model formalizes a \emph{pseudo-variable} as
either a \emph{variable} (an ACL2 symbol)
or the number 1.
A linear combination is formalized as a (sparse) \emph{vector},
i.e.\ an ACL2 list of pairs (ACL2 lists of length 2)
where each pair consists of
a \emph{coefficient} (a field element)
and a pseudo-variable;
the notion of pseudo-variable provides uniformity
between monomials of degrees 1 and 0 in linear combinations.
A \emph{constraint} is formalized as an aggregate
\citeman{STD____DEFAGGREGATE}{defaggregate}
with components \code{a}, \code{b}, and \code{c}
that are the three linear combinations;
this corresponds to the equality $(\code{a})\:(\code{b})=(\code{c})$,
referring to \secref{back}.
Finally, a (\emph{rank-1 constraint}) \emph{system} is formalized as
an aggregate consisting of
a prime, a list of variables, and a list of constraints.
The model also defines well-formedness conditions
on this aggregate and its sub-structures,
e.g.\ that all the variables in the constraints
are also in the list of variables of the R1CS aggregate.
This aggregate and its sub-structures
form the model's formal \emph{syntax} of R1CS.

The \emph{semantics} of R1CS is formalized in terms of
satisfaction of constraints by \emph{valuations},
which are ACL2 alists from variables to field elements,
i.e.\ assignments of field elements to variables.
Given a valuation, a linear combination evaluates to a field element,
in the obvious way;
the model defines this evaluation in terms of \emph{dot product} of vectors,
as in the literature.
Given a valuation, a constraint evaluates to a boolean, in the obvious way.
A valuation \emph{satisfies} a system iff{} it makes all its constraints true.

Besides basic theorems about the syntax and semantics sketched above,
the R1CS library also includes rules for reasoning about R1CS,
for both ACL2 and Axe (see below).

\subsection{Extraction}
\label{extraction}

To verify gadgets using the R1CS model described above,
the gadgets must be represented in the syntactic form defined by the model.
The gadget construction libraries mentioned in \secref{back}
produce R1CS constraints that are not in that form,
and sometimes they do not provide facilities to export them in any form.
Thus, the approach to extract gadgets for verification is case by case.

In a Kestrel project funded by the Ethereum Foundation,
we worked on the verification of Ethereum's Semaphore circuit
\citeman{ZKSEMAPHORE____SEMAPHORE}{semaphore}.
Since Semaphore was written in the high-level language Circom \cite{circom},
whose compiler had a facility to export the R1CS constraints in JSON format,
we developed an ACL2 converter from that format
\citecode
 {books/kestrel/ethereum/semaphore/json-to-r1cs}
 {[books]/kestrel/ethereum/semaphore/json-to-r1cs},
and we used that along with our ACL2 JSON parser
\citecode{books/kestrel/json-parser}{[books]/kestrel/json-parser}.
Taking advantage of the modularity of the Circom code,
we extracted not only the complete circuit gadget,
but also several sub-gadgets.

In a Kestrel project funded by the Tezos Foundation,
we worked on the verification of
Zcash's Jubjub elliptic curve operation circuits
\citeman{ZCASH____ZCASH}{zcash}.
Since these circuits were generated programmatically in Rust,
we instrumented that Rust code, with help from the Zcash team,
to export R1CS constraints directly as s-expressions in the R1CS model's format.
We extracted the top-level gadgets and several sub-gadgets,
by invoking the library at different points.

At Aleo,
we are working on the verification of the snarkVM gadgets \cite{snarkvm-acl2}.
With our Aleo colleagues, we have instrumented snarkVM's Rust code
to export R1CS constraints in JSON format (different from Circom's).
We have developed an ACL2 converter from that format to the model's format,
which we are using along with the ACL2 JSON parser.
We are extracting gadgets at varying levels of granularity.

In the current situation, in all the above cases,
the gadget extraction is trusted:
an error in our instrumentation of the gadget construction libraries,
or in our conversion to ACL2,
may cause us to unwittingly verify
a different gadget from the real one.
However, the top-level gadget $\zkpredlo$
(discussed in \secref{intro}, \secref{back}, and \secref{frame})
is part of the zero-knowledge proof,
which has a well-defined (protocol-dependent) format,
and is generated by tools like snarkVM \cite{snarkvm}
that use or include gadget construction libraries,
That format can be formalized in ACL2,
and the formal verification can be applied directly to
(the $\zkpredlo$ gadget in) the zero-knowledge proof.
In this eventual situation,
the extraction of sub-gadgets of $\zkpredlo$
via instrumentation and conversion
will merely provide building blocks for
the top-level formal proof of $\zkpredlo$
(especially in the compositional approach in \secref{compver}),
but will no longer be trusted.

\subsection{Verification in ACL2}
\label{acl2ver}

Regardless of the exact approach,
the result of the above extraction
is an ACL2 constant, say \code{*gadget*},
whose value is an R1CS aggregate of the form described in \secref{r1cs-model}.
The model confers semantics to this aggregate,
amounting to the relation $\rrel$ in \secref{frame}.
More precisely,
the model provides a predicate
over an assignment of field elements to variables:
\code{(r1cs-holdsp *gadget* asg)}
means that the assignment \code{asg} satisfies
all the constraints in \code{*gadget*},
given the prime that is part of the \code{*gadget*} aggregate
(which is left implicit in $\rrel$).
This can be turned into a finitary relation
over the field elements assigned to the variables,
like $\rrel$,
by specializing \code{r1cs-holdsp}
with an assignment to the gadget's specific variables,
e.g.\ if the variables in \code{*gadget*}
are \code{'x0}, \code{'x1}, \ldots,
the relation $\rrel$ is formalized as
\begin{bcode}
(defun gadget (x0 x1 ...)
  (r1cs-holdsp *gadget* (list (cons 'x0 x0) (cons 'x1 x1) ...))
\end{bcode}
where each \code{'xi} is a variable and each \code{xi} is a field element.

To state and prove correctness,
a specification is written in ACL2,
amounting to $\spec$ in \secref{frame}:
\begin{bcode}
(defun spec (x0 x1 ...) ...)  ; this can be defined in any form
\end{bcode}
If the gadget has no internal variables, correctness is stated as
\begin{bcode}
(defthm gadget-correctness
  (implies (and ...   ; boilerplate hypotheses
                ...)  ; preconditions (if applicable)
           (equal (gadget x0 x1 ...) (spec x0 x1 ...))))  ; R = S
\end{bcode}
where the boilerplate hypotheses say that \code{x0}, \code{x1}, \ldots
are field elements,
and where examples of preconditions are
that some \code{xi} is boolean
or that some \code{xi} is non-zero.
If the gadget has internal variables,
\code{spec} has fewer parameters,
but soundness can be stated and proved similarly,
using \code{implies} in place of \code{equal}.
Completeness is less straightforward;
it is discussed, in a more general context, in \secref{compver}.

The proofs are carried out by
first enabling certain functions of the R1CS semantics,
so that the (evaluated) constraints \emph{deeply embedded} in ACL2
are rewritten to ACL2 terms involving prime field operations,
i.e.\ constraints \emph{shallowly embedded} in ACL2.
Then the core of the proof is handled
via other hints and lemmas, of varying complexity,
that depend on the details of the constraints and specification.

After verifying, in the manner just described,
the correctness of a number of Semaphore sub-gadgets
for elliptic curve operations and data multiplexing
\citecode
 {books/kestrel/ethereum/semaphore}
 {[books]/kestrel/ethereum/semaphore},
two related issues became apparent.
One issue was that the numbers of variables and constraints
and the resulting ACL2 terms
grew quickly as we moved from simpler to more complex gadgets,
making the proofs harder and less efficient.
Another issue was that
because each gadget was extracted in isolation,
with its own specific variable names
generated by the gadget construction libraries
(typically via monotonically increasing indices),
it was not easy to use proofs of sub-gadgets in proofs of super-gadgets:
the same sub-gadget could appear with different variable names
in different super-gadgets, or in different instantiations within the same super-gadget,
but the proof for the separate sub-gadget used different variable names than
would be seen in any of these instantiations.

\subsection{Verification in Axe}
\label{axever}

To combat the growth of terms mentioned in \secref{acl2ver},
we turned to the Axe toolkit \citeman{ACL2____AXE}{axe}.
The \emph{Axe Rewriter} is functionally similar to the ACL2 rewriter,
but it represents terms as directed acyclic graphs (DAGs) instead of trees:
these DAGs share sub-terms,
affording the practical handling of very large terms,
such as fully unrolled AES implementations.

We developed a specialization of the \emph{Axe Lifter} for R1CS,
which turns deeply embedded constraints into shallowly embedded ones,
similarly to what is described in \secref{acl2ver},
and also performs some simplifications of the lifted constraints
using the Axe Rewriter.
This specialized lifter \citeman{R1CS____LIFT-R1CS}{lift-r1cs}
generates an ACL2 constant whose value is a DAG
representing the simplified lifted constraints.

We developed a specialization of the \emph{Axe Prover} for R1CS,
which, given a DAG from the lifter as above
and a specification like \code{spec} in \secref{acl2ver},
attempts to prove soundness \citeman{R1CS____VERIFY-R1CS}{verify-r1cs}
or completeness (via a more general event macro to prove implications).
This specialized prover uses
rewriting and variable elimination via substitution, and it
supports applying different sets of rewrite rules in sequence.
Substitution is enabled by the fact that certain constraints
essentially equate certain variables to expressions over other variables,
though rewriting must often be performed first to make this explicit by solving the constraints.
A constraint is a candidate for substitution if it equates a variable with some sub-DAG not involving that variable.
Large R1CS proofs can involve hundreds or thousands of substitution steps,
and we optimized Axe to apply many substitutions at once when
possible.  For each round of substitution, Axe substitutes a set of
variables each of which is equated to a sub-DAG not involving any
variables in the set.  The set of equalities used in the round is then
removed from the assumptions of the proof.
Repeated substitution of intermediate variables can incrementally turn
a large unstructured conjunction of constraints into a deeply nested
operator tree (represented in DAG form), of the kind commonly verified by Axe.
The ability to apply the rewriting tactic with different sets of rewrite rules
supports the staging of inter-dependent proof steps,
which depend on previous steps and enable subsequent steps.
Suitable rewrite rules can recognize R1CS idioms
and turn them into equivalent higher-level formulations
that may facilitate the rest of the proof.
Similarly,
certain sub-gadgets may also be recognized and raised in abstraction
using rewrite rules based on the correctness properties of such sub-gadgets;
this partially addresses
the second issue described in \secref{acl2ver}.

The Axe verification of the soundness of an R1CS gadget looks like
\begin{bcode}
(lift-r1cs *gadget-dag*  ; name of the generated defconst
           '(x0 x1 ...)  ; variables of the gadget
           ...           ; constraints of the gadget
           ...           ; prime of the gadget
           ...)          ; options
(verify-r1cs *gadget-dag*       ; gadget (simplified and lifted, in DAG form)
             (spec x0 x1 ...)   ; specification
             :tactic ...        ; proof tactics, e.g. (:rep :rewrite :subst)
             ...)               ; other information and options
\end{bcode}

We used this approach
to verify the soundness, and in some cases also the completeness,
of a number of Semaphore and Zcash (see \secref{extraction}) sub-gadgets
that perform fixed-size integer operations,
elliptic curve operations,
instances of the MiMC cipher,
and parts of the BLAKE2s hash
\citecode
 {books/kestrel/ethereum/semaphore}
 {[books]/kestrel/ethereum/semaphore}
\citecode
 {books/kestrel/zcash/gadgets}
 {[books]/kestrel/zcash/gadgets};
these range from relatively small and simple to relatively large and complex.
We also verified the soundness of the large and complex BLAKE2s hash gadget
generated by (an earlier version of) snarkVM \cite{snarkvm};
this is currently not open-source, but it will be in the future.

While using Axe helps address the term growth problem,
the sub-gadget proof re-use problem remained largely unsolved.
The recognition and rewriting of sub-gadgets mentioned above,
which worked for certain cases,
in general may need to recover the sub-gadgets from a sea of constraints.
Each sub-gadget may consist of multiple constraints,
some of which may even have the same form across different sub-gadgets,
requiring the exploration of multiple recovery paths.
Furthermore, constraint optimizations,
such as the ones performed by the Circom compiler and by snarkVM,
which blend gadgets under certain conditions,
may greatly complicate, or defeat altogether, the recovery of sub-gadgets.
Solving these problems is not necessarily impossible, but it is challenging;
as a data point, the aforementioned soundness verification of snarkVM's BLAKE2s
took several person-days to develop and takes several machine-hours to run.

\subsection{Compositional Verification}
\label{compver}

As mentioned in \secref{back},
the hierarchical structure in the gadget construction libraries
gets flattened away in the generated R1CS constraints.
Thus, as discussed in \secrefII{acl2ver}{axever},
the gadgets extracted from the libraries are verified as wholes,
with limited ability to discern their hierarchical structure
and leverage proofs of their sub-gadgets,
resulting in difficult and slow proofs.

More scalability can be achieved
via \emph{compositional verification},
where the proof of a gadget
uses the proofs of its sub-gadgets
and is used in the proofs of its super-gadgets.
This could be accomplished
by extending the gadget construction libraries
to generate such compositional proofs along with the gadgets,
but doing so is impractical
due to the libraries' complexity and ownership.
A viable approach is to
(1) replicate the gadget constructions in the theorem prover,
(2) verify correctness properties of the constructions, and
(3) validate the replicated gadget constructions
by checking that the constructed gadgets are the same as
the ones extracted from the libraries.
We propound the term \emph{detached proof-generating extension}
for this kind of solution.

The gadget constructions are formalized by ACL2 functions
that take variable names as inputs
and return lists of R1CS constraints as outputs.
The constraints are built
either directly or by calling functions that build sub-gadgets,
concatenating all the resulting constraints together.
These functions return lists of constraints,
which are readily composable by concatenation;
they do not return R1CS aggregates (see \secref{r1cs-model}),
which are not readily composable.
The gadget hierarchy corresponds to the function hierarchy.
The parameterization over the variable names is critical,
because separate instances of the same gadget have different variables,
as mentioned in \secrefII{back}{acl2ver}.

To \emph{validate} that these constructions are consistent with the libraries,
we extract \emph{sample gadgets} from the libraries as in \secref{extraction},
and we formulate ACL2 ground theorems saying that
the extracted R1CS constraints are identical to
the ones built by the ACL2 functions
when passed suitable variable names as arguments.
Currently this validation process amounts to
testing our constructions against the libraries.
Eventually, this validation will be performed
every time the libraries are run to generate a zero-knowledge proof,
as explained in \secref{future}.

The correctness of the ACL2 gadget constructions
is proved for generic variable names and generic prime $p$
(sometimes under restricting hypotheses).
The proof opens the function definition
and uses the theorems for any called functions,
whose definitions are unopened;
if the function builds some constraints directly,
certain semantic functions of the R1CS model are also opened,
lifting those constraints to equalities and prime field operations.
Given this proof setup,
the correctness of the gadget (family) built by the function
is proved by reasoning over
the specifications of the sub-gadgets (not the sub-gadgets' constraints)
and/or the constraints of the gadget;
the details depend on the gadget,
and may involve hints and lemmas of varying complexity.

For example, a gadget to force a variable to be boolean as in \secref{back}
is constructed as
\begin{bcode}
(defun boolean-assert-gadget (x)
  (list (make-r1cs-constraint :a (list (list 1 x))              ; (x)
                              :b (list (list 1 1) (list -1 x))  ; (1 - x)
                              :c nil)))                         ; (0)
\end{bcode}
where \code{x} is the variable name to use.
Correctness (soundness and completeness) is expressed as
\begin{bcode}
(defthm boolean-assert-gadget-correctness
  (implies ...  ; boilerplate hypotheses
           (equal (r1cs-constraints-holdp (boolean-assert-gadget x) asg p)  ; R
                  (bitp (lookup-equal x asg)))))                            ; S
\end{bcode}
where \code{asg} assigns field elements to variables,
\code{lookup-equal} retrieves them,
and \code{bitp} is the specification of this gadget;
in the notation of \secref{frame},
this theorem rewrites $\rrel$ ($ = \rrelx$ in this case) to $\spec$.

As another example,
the gadget in \figref{fig:r1cs-eq} is constructed as
\begin{bcode}
(defun equality-test-gadget (u v w s)
  (append (list (make-r1cs-constraint ...))    ; (u - v) (s) = (1 - w)
          (list (make-r1cs-constraint ...))))  ; (u - v) (w) = (0)
\end{bcode}
Soundness is expressed as
\begin{bcode}
(defthm equality-test-gadget-soundness
  (implies (and ...  ; boilerplate hypotheses
                (r1cs-constraints-holdp (equality-test-gadget u v w s) asg p))  ; R
           (equal (lookup-equal w asg)                                          ; S
                  (if (equal (lookup-equal u asg) (lookup-equal v asg)) 1 0))))
\end{bcode}
where the specification of this gadget is that
the value of \code{w} is 1 or 0
based on whether the values of \code{u} and \code{v} are equal or not;
in the notation of \secref{frame},
this theorem derives $\spec$ from $\rrel$ ($ \neq \rrelx$ in this case).

The gadget described in \secref{back} as
the combination of \figref{fig:r1cs-eq} and \figref{fig:r1cs-cond}
is constructed as
\begin{bcode}
(defun if-equal-then-else-gadget (u v x y z w s)
  (append (if-then-else-gadget w x y z)
          (equality-test-gadget u v w s)))
\end{bcode}
which calls the functions for the sub-gadgets
(the definition of \code{if-then-else-gadget} is not shown).

To exemplify varying numbers of variables and constraints,
the gadget in \figref{fig:r1cs-add} is constructed as
\begin{bcode}
(defun addition-gadget (xs ys zs)
  ...  ; guard requires (len xs) = (len ys) = (len zs) - 1
  (append (boolean-assert-list-gadget zs)
          (list (make-r1cs-constraint
                 :a (append (pow2sum-vector xs) (pow2sum-vector ys))
                 :b (list (list 1 1))
                 :c (pow2sum-vector zs)))))
\end{bcode}
where \code{xs}, \code{ys}, and \code{zs} are lists of variables,
\code{boolean-assert-list-gadget}
constructs boolean constraints for all the variables in \code{zs},
and \code{pow2sum-vector} constructs a powers-of-two weighted sum.
The parameterization covers not only the names of the variables,
but also the number of bits $n$ in \figref{fig:r1cs-add},
which is the length of \code{xs} and \code{ys}.
Correctness is expressed as
\begin{bcode}
(defthm addition-gadget-correctness
  (implies (and ...                                      ; boilerplate hypotheses
                (< (1+ (len xs)) (integer-length p))     ; restriction on n
                (bit-listp (lookup-equal-list xs asg))   ; precondition
                (bit-listp (lookup-equal-list ys asg)))  ; precondition
           (equal (r1cs-constraints-holdp (addition-gadget xs ys zs) asg p)
                  (and (bit-listp (lookup-equal-list zs asg))
                       (equal (lebits=>nat (lookup-equal-list zs asg))
                              (+ (lebits=>nat (lookup-equal-list xs asg))
                                 (lebits=>nat (lookup-equal-list ys asg))))))))
\end{bcode}
where \code{lebits=>nat} turns a list of bits into
the integer they denote in little endian order,
and where the restriction on $n$ ensures that the modular weighted sums
can be turned into non-modular sums.
This is proved for every $n$,
using a property of \code{pow2sum} proved by induction.
While the proofs for the previously exemplified gadgets are straightforward,
this gadget takes a little more work.

The details of the examples above,
and of the other ones in \secref{back},
are in the supporting materials, in
\citecode
 {books/workshops/2023/coglio-mccarthy-smith}
 {[books]/workshops/2023/coglio-mccarthy-smith}.
Other examples are in the R1CS library, in
\citecode
 {books/kestrel/crypto/r1cs/sparse/gadgets}
 {[books]kestrel/crypto/r1cs/sparse/gadgets},
where in particular the proofs in \code{range-check.lisp} were quite laborious.

We have employed this approach to verify compositionally
a substantial portion of the snarkVM gadgets \cite{snarkvm-acl2},
specifically most of the ones for boolean, field, and integer operations.
In the process,
we have discovered two bugs in the gadgets, which have been fixed:%
\footnote{The Aleo blockchain mainnet had not been launched yet,
so these bugs did not affect real applications and assets.}
(i) the gadget to convert a field element into its bits
failed to constrain the integer value of the bits to be below the prime,
leading to indeterminacy
(e.g.\ the field element 0 could be converted to
not only all zero bits as expected,
but also to the bits that form the prime,
since $\modulop{p} = 0$);
and (ii) the gadget to calculate square root
allowed both positive and negative roots
(when the input is a non-zero square),
leading to indeterminacy.
We have also identified some possible optimizations,
which have been or are being applied,
saving a large number of constraints in some cases.
Our ACL2 work on snarkVM
is currently not open-source, but it will be in the future.

But even this approach eventually runs into a scalability issue,
due to the internal variables of gadgets.
The names of these variables are exposed as function parameters of
not only the gadgets that directly use them to build constraints,
but also any super-gadgets that contain
(possibly many instances of) those sub-gadgets.
As increasingly large gadgets are constructed,
the function parameters for variable names keep growing,
including all the internal variables at every level.
Furthermore, while soundness theorems
like \code{equality-test-gadget-soundness} above
can ignore internal variables in the consequent of the implication,
completeness theorems need to say something about the internal variables.
In the notation of \secref{frame},
a gadget correctness theorem $\rrelx = \spec$ reduces to $\rrel = \spec$
if there are no internal variables,
which is a good rewrite rule,
as in \code{boolean-assert-gadget-correctness} above.
But internal variables cannot be existentially quantified
in the gadget construction functions,
because these functions must return the gadgets given all their variables.
Instead, the specification $\spec$ over the external variables
must be extended to a specification $\spec'$
over all variables,
including the internal ones at every level.
This exposure of internal variables violates modularity
and impedes compositionality.

\section{Prime Field Constraint Systems}
\label{pfcs}

The scaling issue discussed in \secref{compver}
is addressed by \emph{Prime Field Constraint Systems} (\emph{PFCS}),
a formalism introduced by the authors.
PFCS generalizes R1CS in two ways:
(1) constraints can be equalities between any expressions,
built out of variables, constants, additions, and multiplications; and
(2) constraints can be grouped into \emph{named relations with parameters},
and these relations can be used as constraints
with the parameters replaced by argument expressions
(as in function calls).

The first extension is useful to represent
zero-knowledge circuit formalisms different from R1CS,
but is not especially relevant to verifying R1CS gadgets.
The second extension is important
for verifying R1CS and other kinds of gadgets,
because it explicitly captures their hierarchical structure.
A PFCS relation formalizes a \emph{gadget};
the relation's parameters
are the gadget's external variables,
while the other variables in the relation's defining body
are the gadget's internal variables.%
\footnote{PFCS does not distinguish between input and output external variables.
This distinction only matters to the formulation of the specification $\spec$,
which is still always a relation over the external variables,
as is the semantics $\rrelx$ of the gadget.}
PFCS explicitly handles the existential quantification
that takes $\rrel$ to $\rrelx$:
while $\rrel$ is the semantics of a PFCS relation's body,
$\rrelx$ is the semantics of the PFCS relation itself.
The internal variables of a gadget are taken into consideration
when proving the correctness $\rrelx = \spec$ of a gadget,
which involves $\rrel$,
but can be ignored when proving the correctness of super-gadgets
that include that sub-gadget,
whose semantics $\rrelx$ can be rewritten to $\spec$
in proofs for the super-gadgets;
no extended specification $\spec'$ (see end of \secref{compver}) is needed.

Our development and use of PFCS is still somewhat preliminary.
It is overviewed here,
but it will be described in more detail in future publications.
More information is in the PFCS library \citeman{PFCS____PFCS}{pfcs}.

\subsection{Model}
\label{pfcs-model}

\begin{wrapfigure}[8]{r}{3.8in}
\centering
\vspace*{-0.2in}
\[
\begin{array}{rlcl}
\mathrm{names} &
\pfcsNam &
\pfcsEq &
\langle \mbox{letter then letters/digits/underscores} \rangle \\
\mathrm{integers} &
\pfcsInt &
\pfcsEq &
\ldots \pfcsAlt
\pfcsTerm{-2} \pfcsAlt
\pfcsTerm{-1} \pfcsAlt
\pfcsTerm{0} \pfcsAlt
\pfcsTerm{+1} \pfcsAlt
\pfcsTerm{+2} \pfcsAlt
\ldots \\
\mathrm{expressions} &
\pfcsExp &
\pfcsEq &
\pfcsNam\pfcsAlt
\pfcsInt \pfcsAlt
\pfcsExp \: \pfcsTerm{+} \: \pfcsExp \pfcsAlt
\pfcsExp \: \pfcsTerm{*} \: \pfcsExp \\
\mathrm{constraints} &
\pfcsCon &
\pfcsEq &
\pfcsExp \: \pfcsTerm{=} \: \pfcsExp \pfcsAlt
\pfcsNam \: \pfcsTerm{(} \: \pfcsExp^\ast \: \pfcsTerm{)} \\
\mathrm{relations} &
\pfcsRel &
\pfcsEq &
\pfcsNam \:
\pfcsTerm{(} \: \pfcsNam^\ast \: \pfcsTerm{)} \:
\pfcsTerm{\{} \: \pfcsCon^\ast \: \pfcsTerm{\}}
\end{array}
\vspace*{-0.1in}
\]
\vspace*{-0.1in}
\caption{PFCS syntax.}
\label{fig:pfcs-syntax}
\end{wrapfigure}

The \emph{syntax} of PFCS is approximately described by
the grammar in \figref{fig:pfcs-syntax},
consistently with the informal description above.
A relation $\pfcsRel$ consists of
a name $\pfcsNam$,
a sequence of parameters $\pfcsNam^\ast$,
and a defining body that is a sequence of constraints $\pfcsCon^\ast$.
The abstract syntax is formalized via recursive types
\citeman{ACL2____FTY}{fty}.
The concrete syntax is formalized via an ABNF grammar
\citeman{ABNF____ABNF}{abnf}
complemented by some (upcoming) restricting predicates.

\begin{wrapfigure}[10]{r}{2.7in}
\centering
\vspace*{-0.2in}
\[
\inference
 {\pfcsAsg(e_1) = \pfcsAsg(e_2)}
 {\pfcsSat
  [\ ]
  {\pfcsRels}
  {\pfcsAsg}
  {e_1\:\pfcsTerm{=}\:e_2}}
\]
\[
\inference
 {\setIn
   [\:]
   {\pfcsrel \pfcsTerm{(} \pfcsvar_1 \cdots \pfcsvar_n \pfcsTerm{)}
    \pfcsTerm{\{} \pfcscon_1 \cdots \pfcscon_m \pfcsTerm{\}}}
   {\pfcsRels}
  \\
  \setSup
   {\pfcsAsgx}
   {\setFT
     {\maplet{\pfcsvar_1}{\pfcsAsg(\pfcsexp_1)}}
     {\maplet{\pfcsvar_n}{\pfcsAsg(\pfcsexp_n)}}}
  \\
  \logAll
   [\:]
   {\setIn{i}{\setFT{1}{m}}}
   {\pfcsSat[\ ]{\pfcsRels}{\pfcsAsgx}{\pfcscon_i}}}
 {\pfcsSat
   [\ ]
   {\pfcsRels}
   {\pfcsAsg}
   {\pfcsrel \pfcsTerm{(} \pfcsexp_1 \cdots \pfcsexp_n \pfcsTerm{)}}}
\]
\vspace*{-0.1in}
\caption{PFCS semantics.}
\label{fig:pfcs-semantics}
\end{wrapfigure}

The \emph{semantics} of PFCS is approximately described by
the inference rules in \figref{fig:pfcs-semantics},
which inductively define when
an assignment $\pfcsAsg$ (a finite map from variables to field elements)
satisfies a constraint $\pfcscon$
in the context of a set of relations $\pfcsRels$,
written $\pfcsSat{\pfcsRels}{\pfcsAsg}{\pfcscon}$.
The first rule says that $\pfcsAsg$ satisfies
an equality constraint $\pfcsexp_1\:\pfcsTerm{=}\:\pfcsexp_2$
when the evaluations $\pfcsAsg(\pfcsexp_1)$ and $\pfcsAsg(\pfcsexp_2)$
yield the same field element;
$\pfcsAsg$ extends from variables to expressions in the obvious way.
The second rule says that $\pfcsAsg$ satisfies
a relation constraint
$\pfcsrel \pfcsTerm{(} \pfcsexp_1 \cdots \pfcsexp_n \pfcsTerm{)}$
when $\pfcsRels$ includes the relation
$\pfcsrel \pfcsTerm{(} \pfcsvar_1 \cdots \pfcsvar_n \pfcsTerm{)}
 \pfcsTerm{\{} \pfcscon_1 \cdots \pfcscon_m \pfcsTerm{\}}$
and each constraint $\pfcscon_i$ in its body
is satisfied by an assignment $\pfcsAsgx$
that extends the assignment of
each evaluated argument expression $\pfcsAsg(\pfcsexp_j)$
to the corresponding parameter $\pfcsvar_j$ of the relation;
besides the parameters, $\pfcsAsgx$ must assign field elements
to the other variables (if any) in the body of the relation,
which are internal to the gadget.
Since $\pfcsAsgx$ appears in the premises but not in the conclusion,
it is existentially quantified;
since the values of the relation's parameters are prescribed by the rule,
the existential quantification reduces to
the values assigned to the internal variables (if any),
capturing exactly the existential quantification in $\rrelx$.
The prime $p$ is left implicit in \figref{fig:pfcs-semantics}.

Since ACL2 disallows mutually recursive \code{defun} and \code{defun-sk},
the PFCS semantics is formalized, over the PFCS abstract syntax, via
(1) proof trees for the inference rules
in \figref{fig:pfcs-semantics} and
(2) a proof checker for those proof trees;
that is, a mini-logic is formalized in ACL2.
Since this definition is inconvenient for reasoning about gadgets,
ACL2 rules are provided that capture the inference rules more directly,
without proof trees and proof checker,
as if \code{defun} and \code{defun-sk} were mutually recursive.

\subsection{Verification}
\label{pfcsver}

In the PFCS framework,
gadget constructions are formalized by ACL2 functions
that take no or few inputs
and return (abstract syntax of) PFCS relations as outputs.
Gadgets with fixed numbers of variables and constraints
are built by ACL2 functions with no inputs.
Gadgets with varying numbers of variables or constraints
are built by ACL2 functions whose inputs are
non-negative integers that specify those varying numbers.
None of these ACL2 functions take variable names as inputs,
because variables in PFCS relations are local to the relations
and can be fixed for each gadget:
the external variables, i.e.\ the parameters,
can be replaced when the relations are called;
and the internal variables are existentially quantified.
These ACL2 functions do not call each other,
unlike the ones that construct R1CS gadgets;
the gadget hierarchy is captured directly in the PFCS relations.

Correctness is proved for generic prime $p$
and (if applicable) for generic numbers of variables and constraints
(sometimes under restricting hypotheses).
The \emph{deeply embedded} PFCS relations built by the ACL2 functions
are lifted to \emph{shallowly embedded} PFCS relations,
which are ACL2 predicates over field elements,
with parameters for the external variables
and an existential quantification (via \code{defun-sk})
for the internal variables.
These predicates are defined as conjunctions of
(1) calls of other predicates, one per sub-gadget, and
(2) equalities between terms involving prime field operations,
one per equality constraint;
the predicates' call graph corresponds to the gadget hierarchy.
For gadgets with fixed numbers of variables and constraints,
a deep-to-shallow \emph{lifter} automatically generates the predicates,
along with theorems connecting the deep and shallow formulations;
for gadgets with varying numbers of variables and constraints,
currently the predicate and theorem are manually generated,
but a future extension of the lifter may automate these as well.
Correctness of a shallowly embedded PFCS relation
is proved by opening the predicate definition,
using the called predicates' correctness theorems as rewrite rules,
and using other hints and lemmas of varying complexity as needed.
Correctness is extended to the deeply embedded PFCS relation
via the lifting theorem, in a way that may be automated in the future.

For example,
a PFCS version of \code{boolean-assert-gadget} in \secref{compver}
is constructed as
\begin{bcode}
(defun boolean-assert-gadget ()  ; deeply embedded PFCS relation
  (pfdef "boolean_assert"        ; name
         (list "x")              ; parameter
         (pf= (pf* (pfvar "x")                        ; (x)
                   (pf+ (pfconst 1) (pfmon -1 "x")))  ; (1 - x)
              (pfconst 0))))                          ; (0)
\end{bcode}
The lifter call
\code{(lift (boolean-assert-gadget))}
generates the predicate
\begin{bcode}
(defun boolean-assert (x p)  ; shallowly embedded PFCS relation
  (and (equal (mul x (add (mod 1 p) (mul (mod -1 p) x p) p) p)
              (mod 0 p))))
\end{bcode}
and the lifting theorem
\begin{bcode}
(defruled definition-satp-of-boolean-assert-to-shallow
  (implies ... ; boilerplate hypotheses
           (equal (definition-satp "boolean_assert" defs (list x) p)  ; deep
                  (boolean-assert x p))))                             ; shallow
\end{bcode}
where \code{(definition-satp }$\pfcsrel$\code{ }$\pfcsRels$\code{ (list }%
$\pfcsVal_1$\code{ }$\cdots$\code{ }$\pfcsVal_n$\code{) }$p$\code{)}
formalizes
$\pfcsSat
  {\pfcsRels}
  {\setFT
    {\maplet{\pfcsvar_1}{\pfcsVal_1}}
    {\maplet{\pfcsvar_n}{\pfcsVal_n}}}
  {\pfcsrel \pfcsTerm{(} \pfcsvar_1 \cdots \pfcsvar_n \pfcsTerm{)}}$.
The correctness of the predicate is expressed as
\begin{bcode}
(defthm boolean-assert-correctness
  (implies ...  ; boilerplate hypotheses
           (equal (boolean-assert x p)  ; R (shallow)
                  (bitp x))))           ; S
\end{bcode}
which is extended to the gadget via the lifting theorem as
\begin{bcode}
(defthm boolean-assert-gadget-correctness
  (implies ...  ; boilerplate hypotheses
           (equal (definition-satp "boolean_assert" defs (list x) p)  ; R (deep)
                  (bitp x))))                                         ; S
\end{bcode}

As another example,
a PFCS version of \code{if-equal-then-else-gadget} in \secref{compver}
is built as
\begin{bcode}
(defun if-equal-then-else-gadget ()
  (pfdef "if_equal_then_else"
         (list "u" "v" "x" "y" "z")
         (pfcall "if_then_else" (pfvar "w") (pfvar "x") (pfvar "y") (pfvar "z"))
         (pfcall "equality_test" (pfvar "u") (pfvar "v") (pfvar "w"))))
\end{bcode}
The lifter generates the predicate
\begin{bcode}
(defun-sk if-equal-then-else (u v x y z p)
  (exists (w)
          (and (fep w p)
               (and (if-then-else w x y z p)
                    (equality-test u v w p)))))
\end{bcode}
which existentially quantifies \code{w}
and which calls the lifted predicates for its sub-gadgets (not shown here).
Correctness is expressed as
\begin{bcode}
(defthm if-equal-then-else-gadget-correctness
  (implies ...  ; boilerplate hypotheses
           (equal (definition-satp "if_equal_then_else" defs (list u v x y z) p)  ; R
                  (equal z (if (equal u v) x y)))))                               ; S
\end{bcode}
which rewrites $\rrel$ to $\spec$ without involving the internal variable $w$.

To exemplify varying numbers of variables and constraints,
a PFCS version of \code{boolean-assert\-list-gadget}
mentioned (but not shown) in \secref{compver}
is constructed as
\begin{bcode}
(defun boolean-assert-list-gadget (n)
  (pfdef (iname "boolean_assert_list" n)  ; "boolean_assert_list_<n>"
         (iname-list "x" n)               ; (list "x_0" "x_1" ...)
         (boolean-assert-list-gadget-aux  ; ((pfcall "boolean_assert" (pfvar "x_0"))
          (iname-list "x" n)))))          ;  (pfcall "boolean_assert" (pfvar "x_1"))
                                          ;  ...)
(defun boolean-assert-list-gadget-aux (vars)
  (cond ((endp vars) nil)
        (t (cons (pfcall "boolean_assert" (pfvar (car vars)))
                 (boolean-assert-list-gadget-aux (cdr vars))))))
\end{bcode}
where \code{iname} constructs an indexed name,
\code{iname-list} constructs a list of indexed names,
and the auxiliary function constructs
a list of PFCS relations calls for generic variable names
(which is useful for induction),
which the main function instantiates to specific variable names.
Correctness is expressed as
\begin{bcode}
(defthm boolean-assert-list-gadget-correctness
  (implies ...  ; boilerplate hypotheses
           (equal (definition-satp "boolean_assert_list" defs xs p)  ; R
                  (bit-listp xs))))                                  ; S
\end{bcode}

The details of the examples above,
and of the other ones in \secref{back},
are in the supporting materials, in
\citecode
 {books/workshops/2023/coglio-mccarthy-smith}
 {[books]/workshops/2023/coglio-mccarthy-smith}.
Other examples are in the PFCS library, in
\citecode
 {books/kestrel/crypto/pfcs/examples.lisp}
 {[books]kestrel/crypto/pfcs/examples.lisp}.

We are porting the verified snarkVM gadgets mentioned in \secref{compver}
from R1CS form to PFCS form,
which we will also use for the remaining snarkVM gadgets.

\subsection{Validation}
\label{pfcsvalid}

The PFCS gadget constructions in ACL2
are built in the same way as the R1CS gadget constructions in \secref{compver},
namely by replicating what the gadget construction libraries do.
For the ACL2 R1CS constructions,
different choices of function call graph are possible,
so long as they produce the same R1CS constraints as the libraries.
For the ACL2 PFCS constructions,
different choices of PFCS hierarchy are possible,
so long as, when flattened,
they produce the same R1CS constraints as the libraries.

We plan to develop a \emph{flattener} of PFCS to R1CS,
which will also generate theorems of correct flattening,
i.e.\ that the flattened R1CS constraints
are equivalent to the PFCS constraints.
The flattener will inline all the relation constraints,
resulting in a sequence of equalities,
all of which have the R1CS form
because our PFCS constructions use equality constraints of the R1CS form.

The PFCS gadget constructions in ACL2
will be validated against sample gadgets from the libraries
in the same way as explained in \secref{compver}
for the R1CS gadget constructions in ACL2,
with the addition of the aforementioned PFCS flattener.

\section{Related Work}
\label{related}

The authors are not aware of any other work
to formally verify zero-knowledge circuits using ACL2;
the paper \cite{snarkvm-acl2}
describes our snarkVM verification work in more detail.
There is work using other tools, discussed below.

The $\mathsf{QED}^2$ tool \cite{underconstrained}
is a specialized verifier that combines
a dedicated algorithm with an SMT solver
to automatically establish whether
the outputs of a zero-knowledge circuit
are uniquely determined by the inputs,
or are instead under-constrained;
it may also fail to find an answer.
Their approach is automated,
but our work addresses a stronger property (correctness);
the unique determination of outputs from inputs
is implied by soundness,
when the specification of a gadget is that
the gadget represents a computation
(see \secref{frame}).
Their approach works on individual circuits
like the ones in \secrefII{acl2ver}{axever},
not on parameterized circuit families
like the ones in \secrefII{compver}{pfcsver}.

The SMT solver for finite fields described in \cite{smt-ffields}
has been used to verify automatically
whether circuits produced by certain compilers
are sound
(with respect to the compilation source)
and deterministic
(i.e.\ the outputs are uniquely determined by the inputs,
as in \cite{underconstrained}).
Since our circuit specifications prescribe computations,
in a way that may be similar to the sources of circuit compilers,
their soundness proofs are analogous to ours
(with determinism implied by soundness, at least in our case, as noted above);
but their work does not cover completeness proofs.
Their approach works on individual circuits,
not on parameterized circuit families
(as also noted above for \cite{underconstrained}).
For example, in our work,
an unsigned $n$-bit integer addition circuit family as in \figref{fig:r1cs-add}
is verified once, quickly, for every possible $n$
(see \secref{compver}),
and can be used to verify correct compilation via a syntactic check;
in contrast, in their work,
instances of that family for different values of $n$
are verified separately, taking increasing resources as $n$ grows.
Another advantage of verifying parameterized circuit families
is that their definitions
are essentially formal models of the circuit construction libraries
and therefore help validate the libraries' design and implementation.
The tradeoff between their and our approach is automation versus generality.

The Ecne tool \cite{ecne} uses a dedicated algorithm to perform
weak verification
(their term to mean that the outputs
are uniquely determined by the inputs)
and witness verification
(their term to mean that the outputs and the internal variables
are uniquely determined by the inputs);
their paper also discusses strong verification
(i.e.\ correctness in our work),
but only as future work.
As already noted,
the determinism of output variables is a consequence of soundness in our work.
The determinism of internal variables is unnecessary for correctness,
but it becomes a consequence of correctness
if the latter is stated with respect to an extended specification $\spec'$
that includes the internal variables (as in \secref{compver})
and prescribes their computation from the inputs;
Ecne's witness verification can be thus addressed with our techniques.

There is work on verifying
the compilation of higher-level languages to zero-knowledge circuits
\cite{pinocchioq,ffield-blast,coda}.
While there is probably overlap with our work,
and thus the opportunity for cross-fertilization,
the purpose is a bit different:
we verify the circuits constructed by existing libraries,
which may be used as compilation targets,
or for more general purposes such as programmatic construction of circuits;
as noted above, our approach also helps validate the libraries.

As a final remark,
the notion of existentially quantified circuits (EQCs) in \cite{circ}
is related to
the existential quantification of internal variables in PFCS.

\section{Future Work}
\label{future}

The main thread of future work
is the continued verification of the snarkVM gadgets at Aleo,
extending and improving the ACL2 PFCS library along the way.
We plan to extend the PFCS lifter to work on parameterized gadgets,
which requires a leap in sophistication
in order to operate on the ACL2 functions that construct those gadgets
rather than on the PFCS abstract syntax produced by the ACL2 functions.
We also plan to build a proof-generating flattener of PFCS to R1CS form,
to enable validation against samples extracted from snarkVM
(see \secref{pfcsvalid}).
To handle the gadget optimizations in snarkVM,
we plan to develop proof-generating PFCS-to-PFCS transformations
that correspond to those optimizations:
these can be composed with the proofs for the vanilla (unoptimized) gadgets
to obtain proofs of the optimized gadgets.
The end goal is to verify all the snarkVM gadgets,
including complex ones for cryptographic operations.
These gadgets are being verified against specifications written in ACL2,
which are not directly exposed to prover and verifier
(cf.\ end of \secref{frame});
they are building blocks for the next steps described below.

After reaching the above goal,
the next verification target is
the snarkVM compiler from Aleo instructions
(the assembly-like language
used to represent program code in the Aleo blockchain)
to R1CS constraints,
which uses the snarkVM gadget constructions to generate the constraints.
The approach will be a detached proof-generating extension
of the snarkVM compiler,
built on the detached proof-generating extension
of the snarkVM gadget constructions;
every time the compiler is run,
its generated constraints will be syntactically compared
to the ones generated in ACL2,
including flattening and optimizations as described above,
to ensure that they are identical
and thus that the proof applies to that exact compilation,
as in a verifying compiler.
The specification for the R1CS constraints generated by the snarkVM compiler
is the source Aleo instructions program,
which software developers can read and understand;
the formal proof relies on
a formalization of Aleo instructions that we are building in ACL2.

The same detached proof-generating extension approach will then be used
for the compilation from Leo \cite{leo}
(a high-level programming language for the Aleo blockchain)
to Aleo instructions,
providing an end-to-end verifying compiler functionality
from Leo to R1CS constraints via Aleo instructions.
The specification for the R1CS constraints
generated by the Leo and snarkVM compilers
is the Leo source program,
which software developers can read and understand;
the formal proof relies on
a formalization of Leo that we are building in ACL2 \cite{leo-paper}.

The roadmap delineated above
is part of our overarching work to apply formal verification
to ideally every aspect of the Aleo blockchain and ecosystem.
The aforementioned formalizations of Aleo instructions and of Leo
have more general value than
their role in the formal verification of the compilation.

Alongside the PFCS-based compositional verification approach,
it would also be interesting to continue exploring
the Axe-based whole-gadget verification approach,
in particular to improve the ability to recover sub-gadgets.
There are tradeoffs between the two approaches:
the first one keeps the proofs more manageable and efficient,
but requires the formalization of the gadget constructions;
the second one does not require that formalization,
but needs to recover some of that structure during the proofs.

It would also be interesting to investigate
the use or specialization of Axe for PFCS-based compositional proofs.
Although PFCS aims at keeping proofs relatively small
via parameterization and composition,
Axe may come handy in case some large proof tasks arise.
Axe's tactics may also be useful for certain proofs regardless of size,
and not only for zero-knowledge circuits.

While interactive theorem proving is needed
to verify parameterized circuit families with efficiency and generality,
automated tools like SMT solvers could be useful for certain proof sub-tasks.
ACL2 already has facilities to interface with automated reasoning tools.

There may be opportunities to partially automate
the replication of the gadget constructions in the theorem prover,
in the detached proof-generating extension approach
(see \secref{compver}).
One avenue is the abstraction and translation of
code in the gadget construction libraries.
Another avenue, suggested by a reviewer, is to leverage
any structure that can be recovered from generated gadgets.

\section{Conclusion}
\label{concl}

Our exploration of the zero-knowledge circuit verification problem
has shed more light into the problem,
created and improved libraries and tools of more general use
(e.g.\ the prime fields library),
and evaluated increasingly sophisticated solution approaches.
The PFCS-based compositional approach is promising,
but completing the verification of the snarkVM gadgets
will provide a more definitive validation.

The inherent restrictions on zero-knowledge circuits
might initially lead to think of their verification
as more tractable than general verification.
Our exploration shows the opposite.
As the size and complexity of the circuits grows,
one eventually hits the ``program verification wall''.
This should not be surprising,
for a formalism that can describe sufficiently general computations.
Although zero-knowledge circuits are not Turing-complete,
and their verification is technically decidable because of their finiteness,
the constraint solution space is so large that their verification is
``practically undecidable''.

Our exploration also confirms
the importance of \emph{structure} in formal verification.
Preserving and leveraging the structure
that is naturally available when the circuits are built
promotes more manageable and efficient proofs,
compared to losing that structure and then attempting to recover it.


\section*{Acknowledgements}

Thanks to the Ethereum Foundation
for funding Kestrel's work on the Semaphore circuit verification.
Thanks to the Tezos Foundation
for funding Kestrel's work on the Zcash circuit verification.
Thanks to the Zcash team
for their help in extracting the Zcash R1CS constraints.
Thanks to our colleagues Pranav Gaddamadugu and Collin Chin at Aleo
for their help in extracting the snarkVM R1CS constraints.
Thanks to the reviewers for comments that led to improvements to the paper.


\bibliographystyle{eptcs}
\bibliography{paper}


\end{document}